\documentclass[10pt,prd,preprintnumbers,floatfix,aps,nofootinbib,showpacs,twocolumn,superscriptaddress,noshowpacs]{revtex4} 
\usepackage{epsfig}
\usepackage{url}
\usepackage{hyperref}

\usepackage[normalem]{ulem} 

\usepackage{latexsym}
\usepackage{epsfig}

\usepackage{amsmath}
\usepackage{amssymb}
\usepackage{wasysym}
\usepackage{graphicx}
\usepackage{verbatim}
\usepackage{enumerate,mdwlist}
\usepackage[toc,page]{appendix}
\usepackage{amsfonts}
\usepackage{tikz} 
\usetikzlibrary{calc}
\usepackage{pgfplots}
\usepackage[export]{adjustbox}

\usepackage{dsfont}
\usepackage[normalem]{ulem}
\usepackage{bm}

\newcommand{\ba}{\begin{eqnarray}}
\newcommand{\ea}{\end{eqnarray}}
\newcommand{\be}{\begin{equation}}
\newcommand{\ee}{\end{equation}}

\newcommand{\veff}{v}
\newcommand{\veffb}{\boldsymbol{v}}

\definecolor{grey}{rgb}{0.4,0.4,0.4}
\definecolor{dullmagenta}{rgb}{0.4,0,0.4}
\definecolor{darkblue}{rgb}{0,0,0.4}
\definecolor{midblue}{rgb}{0,0,0.5}
\definecolor{midred}{rgb}{0.5,0,0}
\definecolor{orange}{rgb}{1,0.5,0}
\definecolor{lightbrown}{rgb}{0.75,0.5,0.25}
\definecolor{tan}{cmyk}{0.14,0.42,0.56,0}
\definecolor{djunglegreen}{cmyk}{0.99,0,0.52,0}
\definecolor{lightgreen}{rgb}{0,1,0}
\definecolor{olivegreen}{cmyk}{0.64,0,0.95,0.40}
\definecolor{midgreen}{rgb}{0.0,0.675,0.0}
\definecolor{darkgreen}{rgb}{0,0.5,0}

\usepackage[normalem]{ulem}

\usepackage[all]{xy} 
\usepackage{amsfonts}

\begin{document} 

\title{
Through the lens of Sgr A$^*$: identifying and resolving strongly lensed 
\\ Continuous Gravitational Waves beyond the Einstein radius}

\author{Stefano Savastano}
\email{stefano.savastano@aei.mpg.de}
\affiliation{Max Planck Institute for Gravitational Physics (Albert Einstein Institute) \\
Am Mühlenberg 1, D-14476 Potsdam-Golm, Germany}

\author{Filippo Vernizzi}
\email{filippo.vernizzi@ipht.fr}
\affiliation{Institut de Physique Th\' eorique, Universit\'e  Paris Saclay CEA, CNRS, 91191 Gif-sur-Yvette, France}

\author{Miguel Zumalac\'arregui}
\email{miguel.zumalacarregui@aei.mpg.de}
\affiliation{Max Planck Institute for Gravitational Physics (Albert Einstein Institute) \\
Am Mühlenberg 1, D-14476 Potsdam-Golm, Germany}

\begin{abstract}
Lensed gravitational waves will offer new means to probe the distribution of matter in the universe, complementary to electromagnetic signals. 
Lensed continuous gravitational waves provide new challenges and opportunities beyond those of transient compact binary coalescence. 
Here we consider continuous gravitational waves emitted by isolated neutron stars and lensed by Sgr A$^*$, the supermassive black hole at the center of our galaxy, a system observable by the next generation of gravitational wave detectors. 
We analyze the signatures of this system in detail, addressing parameter estimation and model selection.
Future detectors can distinguish lensed continuous waves and measure their parameters with precision $\sim 1 - 10\%$ for sources within $2-4$ Einstein radii of Sgr A$^*$, depending on the source distance, thanks to the relative motion of the observer-lens-source system. 
 
The chances of observing strongly-lensed neutron stars increase by one order of magnitude relative to previous estimates, thanks to the possibility of detecting lensed systems at several Einstein radii.
Multiple images can be resolved with an angular accuracy $\sim 10$mas, comparable to the best optical telescopes. Image localization probes deviations from axial symmetry and the existence of companions to Sgr A$^*$ in regions complementary stellar orbits and black hole imaging. 
Our methods and many of our results extend to other lenses (e.g. galactic substructure) and sources (e.g. long-lived inspiralling binaries), rendering lensed continuous gravitational waves into versatile probes of astrophysics and fundamental physics.

\end{abstract}

\date{\today}

\maketitle	

\section{Introduction}
Gravitational lensing has become a powerful tool for astrophysics and cosmology, for example in the search for dark objects, exploration of the universe's large-scale structure and measurement of cosmological parameters \cite{Bartelmann:2010fz,Oguri:2019fix,Liao:2022gde}. 

With the recent rise of gravitational wave (GW) astronomy, the lensing of GWs emitted by coalescing binary black holes (BBHs) and neutron stars (NSs) has become the subject of intense research. The coherence, low frequency and frequency evolution of these sources enables the observation of diffraction \cite{Takahashi:2003ix,Dai:2018enj,Caliskan:2022hbu, Tambalo:2022plm,Tambalo:2022wlm, Savastano:2023spl} and phase  \cite{Dai:2017huk,Ezquiaga:2020gdt,Vijaykumar:2022dlp} effects that are challenging to observe with electromagnetic (EM) waves. 
 While no detection of lensed GWs has yet been made \cite{Hannuksela:2019kle,Dai:2020tpj,LIGOScientific:2021izm,Basak:2021ten}, the increasing rate of GW observations offers a promising future \cite{Ng:2017yiu,Xu:2021bfn}.

Besides transient signals produced by explosive binary coalescences, detectors can observe long-lived signals with 
a slow frequency evolution. Sources of these quasi-monochromatic signals fall into two main categories:
1) stellar-mass binaries well before  coalescence, which will be detectable by LISA \cite{LISA:2017pwj,Sesana:2016ljz,Wagg:2021cst}, and exotic binaries, such as sub-solar primordial BBHs, which can be searched with ground-based detectors \cite{Kalogera:2021bya,Miller:2020kmv};
2) rapidly rotating non-axisymmetric neutron stars  with a quadrupolar deformation, producing continuous GWs (CWs), 
observable by ground-based detectors. 
These signals have been searched for in the LIGO-Virgo-Kagra (LVK) data \cite{LIGOScientific:2022pjk, Steltner:2023cfk,Dergachev:2022lnt}; see Refs.~\cite{Riles:2017evm,Sieniawska_2019,Piccinni_2022,Riles:2022wwz} for reviews.

Isolated spinning neutron stars (NSs) are expected to produce CWs in the $10^2-10^3\,\textrm{Hz}$ band through a variety of different mechanisms \cite{Glampedakis_2018, Sieniawska_2019, Piccinni_2022}, and could be observed by future ground-based detector campaigns. Neglecting the effect of the frequency evolution and the detector's motion and response, the NS signal is monochromatic. For an estimate of the total number of neutron stars that can be probed with current and future detectors, we refer to \cite{Reed:2021scb, Pagliaro:2023bvi}.

The lensing of CWs presents interesting and distinct characteristics. Specifically, due to the coherent nature of CWs, a gravitational lens can act as a diffractive barrier, resulting in interference fringes on the detected signal when the observer-lens-source system undergoes relative transverse motion \cite{Liao:2019aqq,Suvorov:2021uvd,Moylan:2007fi,Jung:2022tzn,Takahashi:2023gdt}. Additionally, CWs allow for very precise sky localization, impossible with binary coalescence observations. Despite the promising prospects for distinguishing between lensed and unlensed CWs, the ability of observations to accurately infer the lens parameters remains an open question.

While no clear detection of CWs has been reported yet, the sheer number of NSs in the Milky Way (MW) galaxy, estimated in $\sim 10^9$ from population synthesis studies \cite{Narayan:1987}, suggests that detection of lensed signals is plausible for future observations. 
Motivated by such a prospect, we study the phenomenology of strongly-lensed CWs, the inference of lens parameters and the potential for detection by next-generation GW observatories. 

For concreteness, we focus on monochromatic, isolated, rotating NS moving at constant velocity and lensed by Sgr A$^*$, the supermassive black hole (BH) at the center of the MW \cite{Ghez:2000ay,Genzel:2010zy,EventHorizonTelescope:2022wkp}. 
In fact, depending on their distribution, a fraction of NSs in our galaxy is expected to lie close to the line of sight of Sgr A$^*$. Moreover, they are expected to have sizeable projected transverse velocities \cite{Sartore:2009wn} due to either natal kicks ($\sim 400$ km/s) \cite{Hobbs:2005yx}, or the Solar System's motion in the galaxy ($\sim 200$ km/s) \cite{Sofue:2008wt,Mr_z_2019, Eilers_2019}. These objects are the subject of targeted CWs searches \cite{LIGOScientific:2022lsr}. Sgr A$^*$ can act as a foreground lens for sources sufficiently aligned with our line of sight. Third-generation (3G) interferometers will be able to observe up to $\sim 6$ strongly lensed signals of rapidly spinning NSs within the Einstein cone of Sgr A$^*$ \cite{Basak:2022fig}, the precise number depending on the NS properties and their spatial distribution in the MW \cite{Reed:2021scb}.

In this paper, we show that CWs lensed by Sgr A$^*$ can be detected even when the source is located outside the Einstein cone so that the expected number of detectable sources increases by an order of magnitude with respect to previous studies. Moreover, we show that it is possible to sky-localize the images from lensed CWs and quantify the uncertainty for future detectors. Finally, we demonstrate the potential use of lensed CWs to explore the structure of the Galactic center. While the Sgr A$^*$ scenario is compelling, our methods and many of our conclusions extend to other systems involving lensed CWs.
The paper is structured as follows. Section \ref{sec:lensing} delves into the phenomenology of gravitational waves lensing and explores its interaction with standard search methods. In Section \ref{sec:observation}, we demonstrate that lens properties can be extracted from the observed signals, even when the source and the lens are located at a few Einstein radii apart. Additionally, we examine the likelihood of such occurrences and highlight the feasibility of resolving individual lensing images.
Lastly, in Section \ref{sec:probing}, we show that by resolving the images, it is possible to probe the existence of additional objects in the vicinity of Sgr A$^*$, as they would cause a misalignment of image positions with the optical axis.


\section{Strong lensing imprints on CWs}\label{sec:lensing}

In this section, we discuss how gravitational lensing alters CWs signals, in the regime of strong lensing. We will first introduce the basics of gravitational lensing (Sec. \ref{sec:lens_theory}). We will then present the features that a moving lens imprints on a CW signal (Sec. \ref{sec:lensing_signatures}). Finally, we will discuss the interplay between lensing and CWs search methodologies (Sec. \ref{sec:seaches}).

\subsection{Gravitational lensing}\label{sec:lens_theory}

Using spherical coordinates centered at the observer, $\vec{r} = (r, \theta, \varphi)$, we define the two-dimensional vector $\boldsymbol{\theta} = \theta (\cos \varphi, \sin \varphi)$. We denote by $\boldsymbol{\theta}_l$ and $\boldsymbol{\theta}_s$ the observed and true angular positions of the source, respectively. Additionally, we use $r_l$, $r_s$, and $r_{ls}$ to represent the observer-lens, observer-source, and lens-source angular-diameter distances, respectively; see Fig.~\ref{fig:interference_pattern}.

In  wave optics, the frequency-domain amplification factor  due to the presence of a lens at rest, $F \equiv \tilde{h}_l/\tilde{h}_0$,
is given in the form of a diffraction integral \cite{NakamuraDeguchi,Takahashi:2003ix},
\begin{equation}
F(\omega, \boldsymbol{\theta}_s)=\frac{\omega}{2 \pi i}\frac{r_l r_s}{ r_{ls}}\int d^{2} \boldsymbol{\theta} \exp [i \omega t(\boldsymbol{\theta}, \boldsymbol{\theta}_s)] \;.
\end{equation}
Here we have defined
\begin{equation}
t(\boldsymbol{\theta}, \boldsymbol{\theta}_s)= \frac{r_l r_s}{2 r_{ls}} |\boldsymbol{\theta}-\boldsymbol{\theta}_s|^{2}- \hat \psi(\boldsymbol{\theta})\,,
\end{equation}
with the lensing potential given as $ \hat \psi (\boldsymbol{\theta}) \equiv \int d r\,  U(r, \boldsymbol{\theta})$.  
In the limit of geometric optics (GO), which applies when $\omega t(\boldsymbol{\theta}, \boldsymbol{\theta}_s) \gg 1$, the diffraction integral is dominated by the stationary points of $t(\boldsymbol{\theta}, \boldsymbol{\theta}_s)$:
\begin{equation}
    F\equiv \frac{\tilde {h}_l(f)}{\tilde {h}_0(f)} =\sum_j \sqrt{|\mu_j|}
    \exp\big[2\pi i\left( f t_j + \pi n_j\right)\big]\,. \label{eq:amplifact}
\end{equation}
Each addendum in the sum corresponds to a GO image, with image position $\boldsymbol{\theta}_{j} $ determined by the lens equation~\cite{Schneider:1992},
\begin{equation}
\label{Fermateq}
\nabla_{\boldsymbol{\theta}} t = 0 \;.
\end{equation}
Moreover, for each image,  $\boldsymbol{\alpha}_j\equiv\frac{r_l}{r_{ls}}(\boldsymbol{\theta}_{j}-\boldsymbol{\theta}_s)$ denotes its deflection angle, $\mu_{j}(\boldsymbol{\theta}_s) \equiv \det(\partial \boldsymbol{\theta}_s/\partial \boldsymbol{\theta}_{j})^{-1}$ its magnification, and $t_j( \boldsymbol{\theta}_s) \equiv t(\boldsymbol{\theta}_{j},\boldsymbol{\theta}_s )$ its time of arrival, which is a solution of the above equation. The Morse phase index, $n_j$, is $0$, $\pi/2$ or $\pi$ depending on whether the image corresponds to a minimum, saddle point or maximum of $t(\boldsymbol{\theta}, \boldsymbol{\theta}_s)$.

\begin{figure}
    \centering
    \includegraphics[width=1\columnwidth]{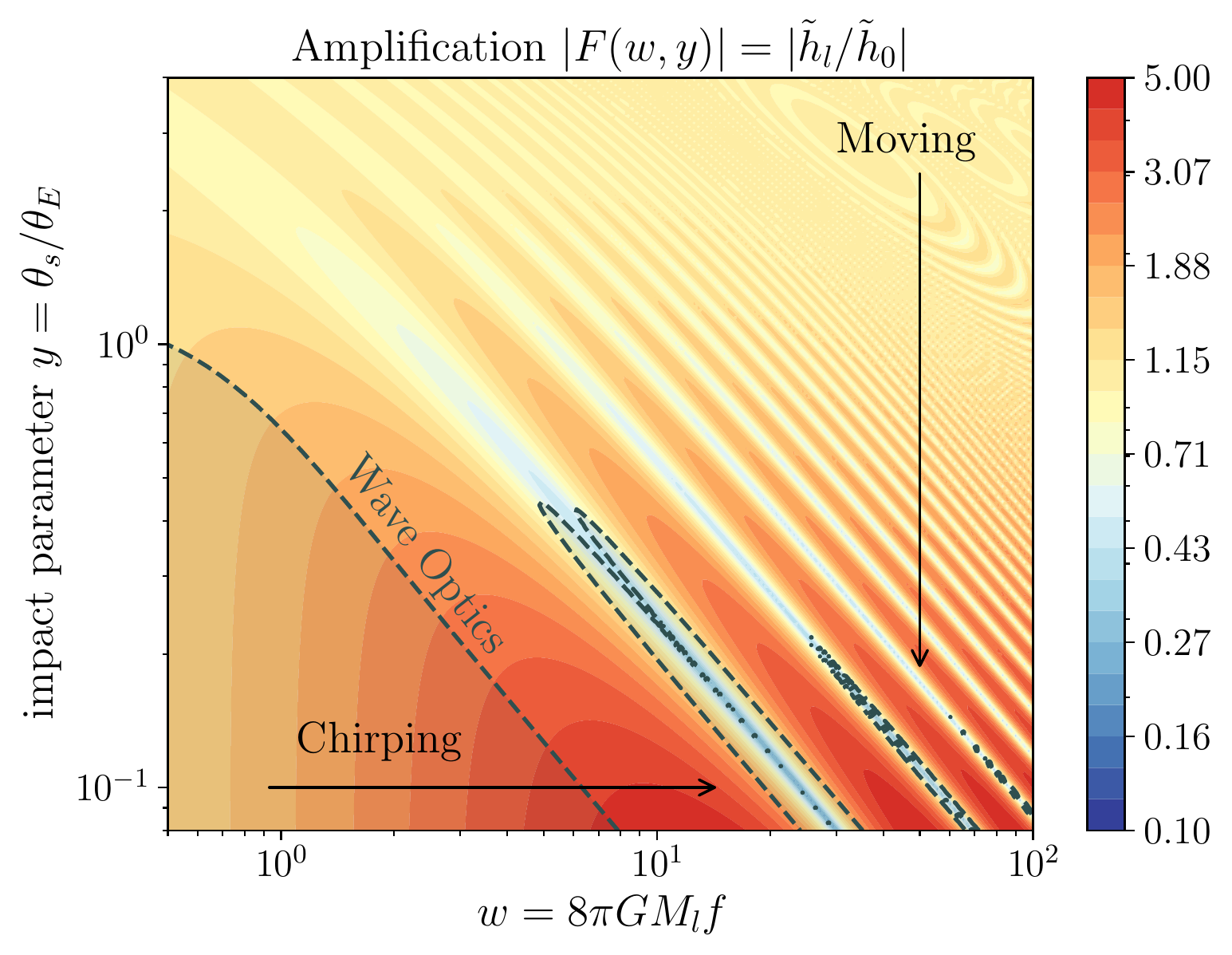}
    \caption{Amplification factor of a point lens as a function of the dimensionless frequency $w = 8 \pi G M_l\, f$ and impact parameter $y = | \boldsymbol{\theta}_s|/\theta_E$. Dashed grey contours correspond to the transition between wave and geometric optics regimes (less than $10\%$ relative difference). Diffraction and wave effects can be observed for chirping binaries (varying $w$ and keeping $y$ constant) or moving lens systems (varying $y$ and keeping $w$ constant) as explored in this work. Our fiducial system lies in geometric optics with $w=8 \pi G M_{\rm Sgr A^*} f_0 \simeq 10^3-10^4$. }
    \label{fig:wave_optics}
\end{figure}
For convenience, we convert angles to dimensionless coordinates, $\boldsymbol{x}$,  by normalizing by the Einstein angle 
\begin{equation}\label{eq:theta_e}
{\theta_E \equiv \sqrt{ \frac{4 G M_l  r_{ls}}{r_l r_s}}}\,,
\end{equation}
namely $\boldsymbol{x}\equiv{\boldsymbol{\theta}}/{\theta_E}$. In particular, we  indicate the $j$-image's position as $\boldsymbol{x}_j\equiv {\boldsymbol{\theta}_j}/{\theta_E}$ and  the dimensionless impact parameter as $\boldsymbol{y} \equiv \boldsymbol{\theta}_s/\theta_E$. We choose the horizontal axis of our coordinate system to coincide with the optical axis, i.e. $\boldsymbol{y}=(y,0)$.
Moreover,  we define the dimensionless frequency, $w\equiv 4 G M_l\,\omega$, where $M_l$ is the lens mass.

Figure \ref{fig:wave_optics} shows $|F|$ as a function of $w$ and $y$ for a point lens and highlights the region where geometric optics applies.
A point-like lens always splits the source into two images, identified by $+$ and $-$, with \cite{Schneider:1992}
\begin{align} \nonumber
    \frac{\mu_{-}}{\mu_{+}}&= \frac{2-y(\sqrt{y^2+4}-y)}{2+y(\sqrt{y^2+4}+y)} \;, \\ \nonumber
    \Delta t_{+-} &= {2 G M_l}\left[y\sqrt{y^2+4}+2 \log \left(\frac{\sqrt{y^2+4}+y}{\sqrt{y^2+4}-y}\right)\right]  \;, \\ \nonumber
    x_\pm &= \frac{1}{2}\left(y\pm \sqrt{y^2+4}\right)\\ \nonumber
    \alpha_{\pm} &= \sqrt{ \frac{4 G M_l  r_s}{r_l r_{ls}}} x_{\pm} \,,
\end{align}
where we denote by $\Delta t_{+-}$ the time delay elapsed between the two images.
 For example, at $y=1$ and $r_{ls}/r_{l}=1$, lensing by Sgr A$^*$ induces a $\Delta t_{+-} =  163.9\,{\rm s}$ and $\Delta\alpha_{+-} =6.3''$.  
We will not consider the effects of external convergence and shear \cite{An:2006bq}, as they will produce negligible corrections for galactic lenses at $y\sim \mathcal{O}(1)$.

\subsection{Lensing signatures in CWs}\label{sec:lensing_signatures}

Lensed CWs can be determined via
\begin{enumerate}
    \item source frequency evolution,
    \item modulation due to transverse motion, or 
    \item spatially resolved images.
\end{enumerate}
\noindent In Fig.\ref{fig:wave_optics} we illustrate the modulation in amplification factor produced in the first two cases. Typically, rapidly rotating NSs  have small period variations $\dot P \lesssim 10^{-18}$s/s \cite{Harding:2013ij}, precluding frequency evolution from revealing lensed systems. Below we discuss the transverse motion modulation, leaving the discussion of spatially resolved images for Sec.\ref{sec:resolve_images}). 

In the presence of transverse motion,  the time delay of an image $j$, $t_j$, can be expanded at linear order in $t$ around a reference time $t_0$, hence 
\be \label{eq:tj}
t_j(t)=t_j(t_0) + \boldsymbol{\alpha}_j(t_0)\cdot \veffb (t- t_0),
\ee
where $\boldsymbol{\alpha}_j\equiv\frac{r_l}{r_{ls}}(\boldsymbol{\theta}_{j}-\boldsymbol{\theta}_s)$ is the deflection angle for the $j$-th image, and $\veffb$ is
 the \textit{projected transverse velocity,} given by \cite{Kayser}
\be
\veffb = \boldsymbol{v}_l - \frac{r_l}{r_s} \boldsymbol{v}_s - \frac{r_{ls}}{r_s} \boldsymbol{v}_o \;,
\ee
where $\boldsymbol{v}_{o}$, $\boldsymbol{v}_{l}$ and $\boldsymbol{v}_{s}$ are the transverse velocity of the observer, lens and source in their respective planes. 
Projected transverse motion results in a time variation of the lensing functions. In fact, this is equivalent to considering a source position that varies with time. Specifically, we have 
\begin{equation}
{y(t)=\sqrt{y_0^2+\left(\frac{\veff (t - t_0)}{r_l \theta_E}\right)^2}}\,,
\end{equation}
where $y_0$ is the impact parameter at $t=t_0$. 
In general, the magnification and deflection angle of each image also acquire a time dependency from the transverse motion, but the effect is negligible for heavy lenses not closely aligned with the sources, as considered here.

\begin{figure} 
    \centering    \hfill\includegraphics[width=.95\columnwidth]{diagram.pdf}  \includegraphics[width=1.\columnwidth]{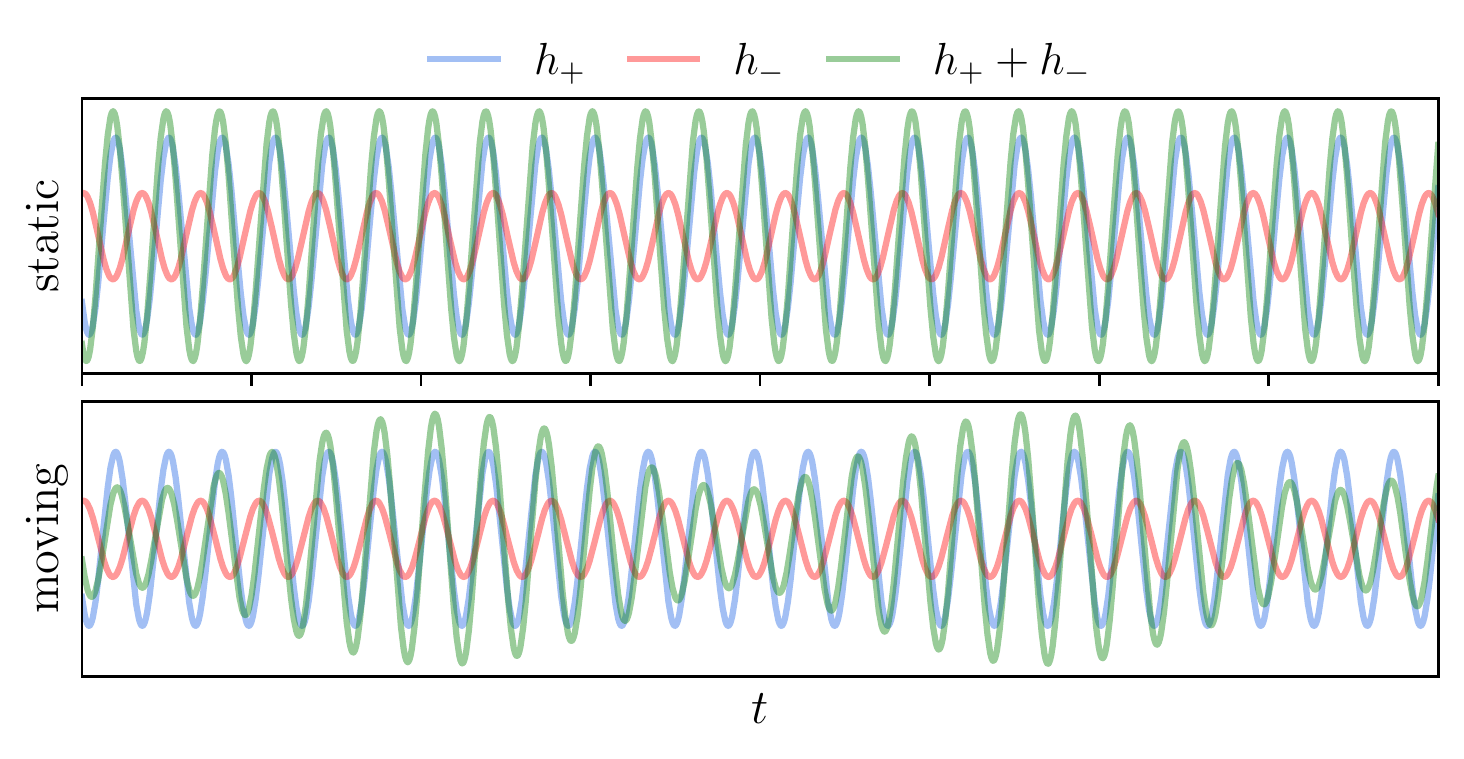}
    \caption{Strong lensing of a quasi-monochromatic source. 
    Top: Gravitational lensing by Sgr A$^*$ forms two images of a rotating NS. Here, $\boldsymbol{\eta}=\boldsymbol{\theta}_s r_s$ is the physical impact parameter in the source plane, related to $y$ by $|\boldsymbol{\eta}|=y \theta_E r_s$. 
    Middle: static case, $h_+,h_-$ are magnified and delayed. Their interference at the detector is again a monochromatic waveform.
    Bottom: moving case,  the images are additionally slightly red/blue-shifted. Their interference produces a modulated signal. (here exaggerated for illustrative purposes). 
    }
\label{fig:interference_pattern}
\end{figure}

The time-varying time delay in Eq.\eqref{eq:tj} induces an effective blue/red-shift on each image's frequency by a 
\begin{equation}\label{eq:freq_sft}
    z_j = \frac{d}{dt}{\ln(t_j)}=\veffb \cdot \boldsymbol{\alpha}_j.
\end{equation}
When two images $i$ and $j$ interfere in the detector, the strain exhibits an amplitude and phase modulation arising from the difference $z_i-z_j$; see Fig.\ref{fig:interference_pattern}. Given a lensing signal with lens mass $M_l$, effective lens velocity $\veff \equiv |\veffb|$, and impact parameter $y_0 \equiv y(t_0)$,
potential information on these parameters is contained in three features: 1) the modulation period,
\begin{equation}\label{eq:modulation_period}
    T_{ij} = \frac{1}{f_0 \veff  \alpha_{ij} }  = 2.4 \, {\rm d} \left(\frac{1 {\rm kHz}}{f_0}\right)
     \left(\frac{10^{-3}}{\veff}\right)
    \left(\frac{1''}{\Delta \alpha_{ij}}\right)\,,
\end{equation}
where $\Delta \alpha_{ij}=|\boldsymbol{\alpha}_i-\boldsymbol{\alpha}_j|$;
2) the modulation amplitude  $\sqrt{\mu_j/\mu_i}$, which depends purely on $y(t)$; and 
3) the modulation phase, given by $\Delta t_{ij}\propto M_l$.

Additional effects on CWs (detector motion, source's frequency evolution or orbital motion, two overlapping sources), 
can be corrected for, or distinguished from, the modulation signature. We discuss them in the next section.


\subsection{Interplay between lensing and CWs searches} \label{sec:seaches}
Now, we will briefly summarize the methodology of CW searches and their interplay with the occurrence of lensed signals in the datastream. The aim of this section is to show that standard searches are able to capture lensed events.

In blind searches the data is pre-processed assuming certain source properties (location, frequency evolution) and then the monochromatic CW signals are searched via a fast-Fourier transform \cite{Brady:1997ji} (we will follow Ref.~\cite{Maggiore:2007ulw}, Ch.~7 \cite{Maggiore:2007ulw}). In particular, the pre-processing is necessary to clean the signal from other sources of distortion such as the orbital motion and frequency evolution of the signal, and search for a monochromatic waveform:
\be
\label{eq:h0QM}
h_0(t, r_s) = \frac{\cal A}{r_s} \,   e^{-i \phi(t)} \;, \qquad \phi (t) = 2 \pi f_0 t+\phi_0   \;,
\ee
where ${\cal A}$ depends on the source's orientation and $\phi_0$ is the phase value at $t=0$.
The signal-to-noise ratio (SNR) of this signal is (see App.\ref{app:parinfer})
\be\label{eq:snr}
{\rm SNR} \simeq \sqrt{ \frac{\mathcal{A}^2 }{r_s^2} \frac{T_{\rm obs}}{ S_n(f_0)} } \,,
\ee, 
where $S_n$ is the one-sided power spectral density and $T_{\rm obs}$ is the observational time. The dominant mode of a CW emitted by a NS has amplitude $ {\rm \cal A}=4 \pi^{2} G {I_{z z} f_{\mathrm{0}}^{2}}\epsilon\,$, where $I_{z z}$ 
is the moment of inertia of a perpendicular biaxial rotor spinning with axis $\hat z$ and the ellipticity parameter, $\epsilon=(I_{x x}-I_{y y}) / I_{z z}$, describes the degree of anisotropy around the NS rotation axis. 
To date, no CWs have been detected but frequency-dependent upper limits on their amplitude and on $\epsilon$ at a fixed distance were inferred \cite{LIGOScientific:2022pjk,LIGOScientific:2022lsr,Dergachev:2022lnt}.  
We adopt $\epsilon=10^{-7}$ and $I_{zz}=$ $10^{38} {\rm\,kg\,m^2}$ \cite{LIGOScientific:2022pjk}, compatible with data, theoretical expectations
\cite{Suvorov2016,Gittins_2020} and the lower bound from population-based studies, $\epsilon \gtrsim 10^{-9}$ \cite{Woan:2018tey}.

This scheme is referred to as \emph{coherent search}. Under this type of search, strongly lensed CWs would first appear as two signals very close in frequency and sky localization.

The large number of unknown parameters and the large observation time required for pre-processing signals make coherent searches computationally infeasible. Instead, searches are typically performed using a semi-coherent method~\cite{Piccinni:2019zub}, in which data streams are divided into segments of duration $T_{\rm coh}$. These segments are processed as described below and then combined incoherently, i.e. neglecting the relative phase between stacks. Therefore, GW detectors are capable of observing NSs up to distances of \cite{Maggiore:2007ulw}
\begin{equation}
\label{eq:NS_horizon}
r_{\rm hor}= 35\,{\rm kpc}\times \left(\frac{\epsilon}{10^{-7}}\right)\left(\frac{f_0}{800\,{\rm Hz}}\right)^2 \left(\frac{T_{\rm obs}}{10\,{\rm yr}}\right)^{1/2} \gamma \;,
\end{equation}
where we have defined an efficiency factor,
\begin{equation}\nonumber
{\gamma \equiv \left(\frac{3750}{\mathcal{N}}\right)^{1/4}}\left(\frac{4}{{\rm SNR}_{\rm thr}}\right)\left(\frac{4.8\,10^{-25}}{\sqrt{S_n(f_0)\, {\rm Hz}}}\right)
\end{equation} 
using the reference numbers expected in next-generation detectors \cite{Maggiore:2007ulw}. 
Here, ${\rm SNR}_{\rm thr}$ is the SNR threshold of the search, 
and ${\mathcal{N}={T_{\rm obs}}/{T_{\rm coh}}}$ is the number of stacks in which the data is divided. The signal's amplitude has been sky-averaged over the solid angle and polarization angle. We stress that once a CW detection from the Galactic center has been confirmed, follow-up analyses can exploit the full coherence of the signal, with no SNR loss. The outcomes of our study hinge on this premise.

The first step in obtaining a monochromatic waveform is a {\em resampling} of the signal to account for the detector's motion. This is done by a redefinition of the time variable: $t=t^{\prime} + \hat n\cdot \vec x(t^\prime) + \Delta T$ \cite{Brady:1997ji}. 
Here, $t^\prime$ is the observation time, $\hat n$ is the source's direction, $\vec x$ is the detector's position relative to the Solar System barycenter and $\Delta T$ is the relativistic time delay. The antenna pattern (given by $\hat n$) has been factored out of ${\cal A}$.
As discussed in Ref.~\cite{Maggiore:2007ulw} (see Eq.~7.151), resampling requires accuracy on the source's position at the level of
$\delta\theta \lesssim 22' \left(1{\rm d}/{T_{\rm coh}}\right)^2\left({\rm kHz}/{f_0}\right)$, with 
$1\, {\rm d} \lesssim T_{\rm coh} \lesssim 10^2 \, {\rm d}$.
This limit corresponds to all-sky NS searches, where a relatively low  $T_{\rm coh}$ allows computational efficiency \cite{Brady:1997ji}. 
Since the separation between images is $|\alpha_{\pm}| \lesssim 10''$ data resampling is common to all images for $T_{\rm coh}\lesssim 11$ d. For larger coherence time, they will show up in two different sky bins and fine sky localization can be achieved (see Sec \ref{sec:resolve_images}).

An additional step is to correct the source's frequency evolution \cite{Brady:1997ji}. This is done by {\em demodulating} the signal via a time coordinate redefinition $t\to t-t_0 + \dot f_0/(2f_0)(t-t_0)^2+\cdots$, ($\ddot f_0$ and higher derivatives can be considered at this stage).
Frequency evolution affects the CW's phase on a timescale $(\dot f_0/2)^{-1/2} = 16.4  \,{\rm d}\left(\frac{1{\rm pHz}/{\rm s}}{\dot f_0}\right)^{1/2}$, making demodulation a necessary step of the analysis. 
However, the interplay between lensing and $\dot f_0$ for rapidly rotating NSs is typically negligible: it can produce additional modulation, due to a phase difference between images on a timescale of  $(\dot f_0 \Delta t_{\pm})^{-1}  = 317 {\rm yr}\left(\frac{100{\rm s}}{\Delta t_{\pm}}\right)\left(\frac{1{\rm pHz}/{\rm s}}{\dot f_0}\right)$. For ms pulsars ($\dot f\lesssim 1$pHz/s) \cite{Woan:2018tey}, this is much longer than $T_{\rm obs}$ and 
treating frequency evolution as common to all images is an excellent approximation. Hence, demodulation does not affect the search of lensed CW or their analysis.

\begin{figure}
    \centering
    \includegraphics[width =\columnwidth]{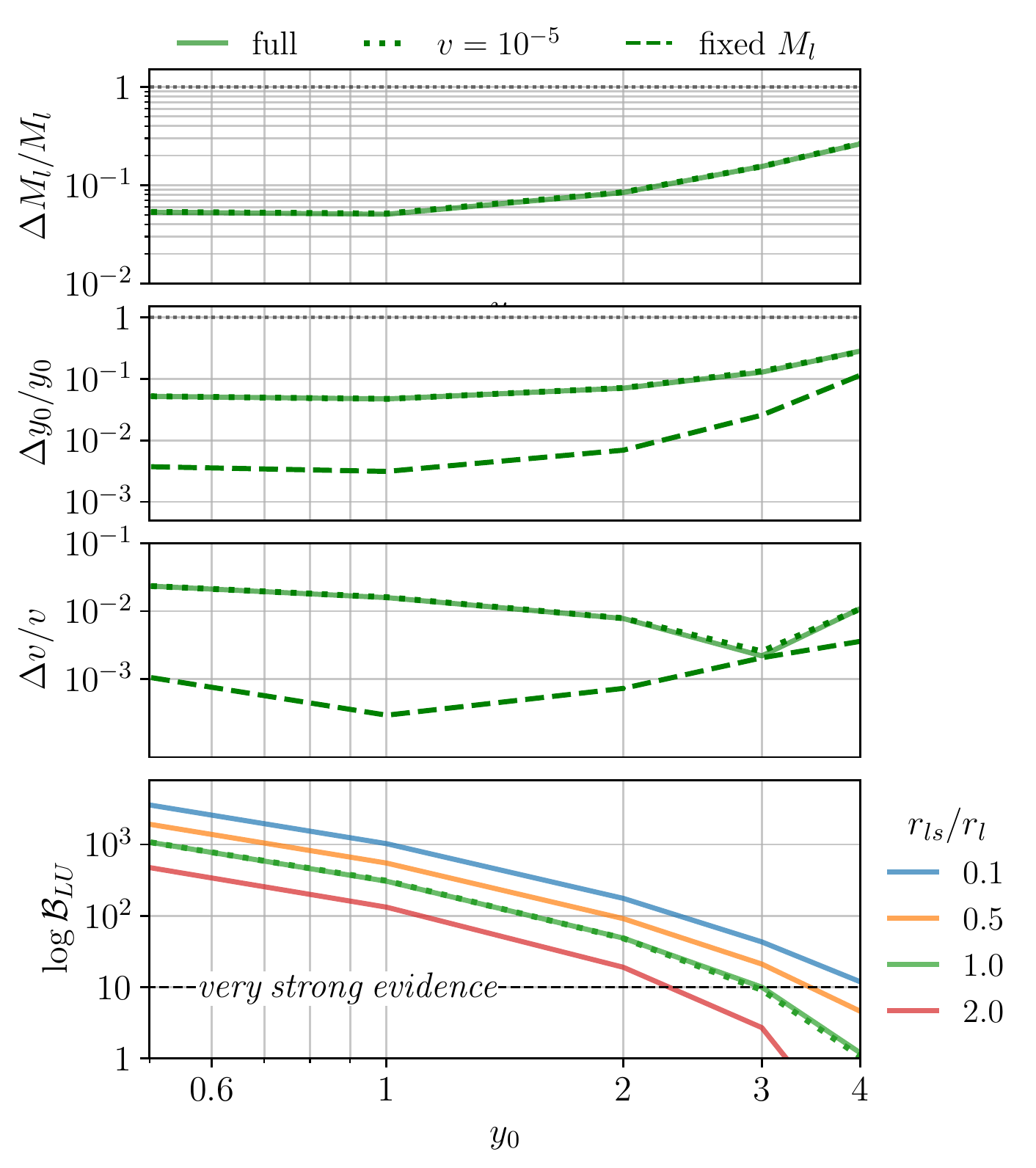}
    \caption{MC results. For the fiducial setup, two cases are considered: uniform prior on the lens mass (solid green line); fixed lens mass (dashed green line). For variable mass, the 
    accuracy remains for very small relative velocity, $\veff=10^{-5}$ (dotted). 
    Top three panels: 68\% C.L.~marginalized error on $M_{l}$ (top), $y_0$ (middle) and $v$ (bottom), as a function of the initial impact parameter $y_0$.  
    Bottom panel: Bayes' ratio of lensed vs unlensed hypothesis against the initial impact parameter for multiple distance ratios, $r_{ls}/r_l$. $\log {\cal B}_{LU} > 10$ 
    corresponds to \emph{very strong evidence} for the lensing hypothesis. The purple line shows the case of advanced LIGO for $r_{ls}/r_l=1$.
    }
    \label{fig:comparison}
\end{figure}

\begin{figure}
    \centering
    \vspace{18pt}
    \includegraphics[width =\columnwidth]{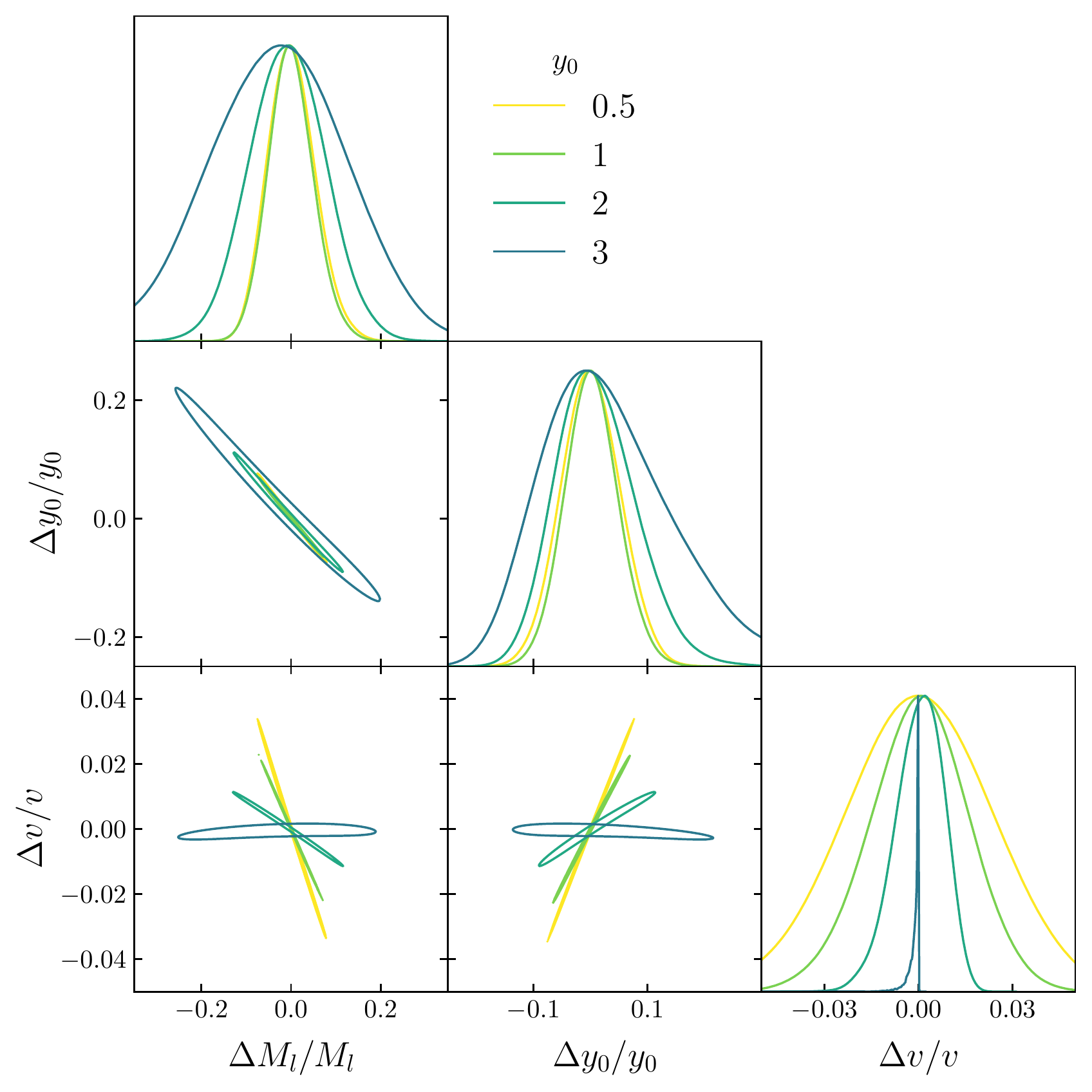}
    \caption{1-d and 2-d marginalized 1-$\sigma$ C.L.~posteriors of the lens parameters, for different initial impact parameters (fiducial setup).}
    \label{fig:triangleplot}
\end{figure}


\section{Observing lensed CWs}\label{sec:observation}
Let us now discuss the prospects for observing and reconstructing lensed CWs. First, we will discuss that lens parameters can be extracted from the observed signal (Sec. \ref{sec:pe}) and that lensing images can be individually resolved and localized (Sec. \ref{sec:resolve_images}). Additionally, we discuss the probability of detecting favourable events (Sec. \ref{sec:probability_estimates}).

\vspace{5pt}
\subsection{Lens parameter reconstruction}\label{sec:pe}
Once a search detects the two images of a lensed CW signal, it is possible to exploit the full-coherent datastream to extract the lens parameters, $M_l$, $y_0$  and $ v$, from the modulation of the lensed signal discussed in Sec. \ref{sec:lensing_signatures}.

We assume a single Einstein Telescope detector and in-band signal duration $T_{\rm obs}= 10\, {\rm yr}$. Parameter estimation will improve using a global network of 3G detectors \cite{Kalogera:2021bya}, with increased SNR and sky coverage. As a fiducial setup, we consider a point-like lens with the measured mass of Sgr~A$^*$, $M_l=4.154\cdot 10^6\,M_\odot$ \cite{Gravity2019} and lens velocity $\veff =10^{-3}$, at a distance $r_l=8.178\, {\rm kpc}$ \cite{Gravity2019}. A point lens produces a brighter and a fainter image, indicated respectively as $+$ and $-$. The source is a spinning NS emitting CWs at a frequency $f_0=800\, {\rm Hz}$  and relative distance $r_{ls}= r_l$.  We will discuss how the results change away from these fiducial values. 

We follow a Bayesian approach and compute their marginalized posteriors through numerical Monte Carlo (MC) sampling. 
In particular, we sample the likelihood function, which follows from the definition in Eq.~\eqref{eq:whittle_likelihood}, through the dynamic nested sampler Dynesty \cite{Speagle_2020}. 
Our analysis is restricted to the source and lens parameters that are potentially correlated. In particular, the source is assumed to be purely monochromatic and modelled by $({{\cal A}/r_s, f_0, \phi_0})$. The original sampled lensing parameters are 
\begin{equation}\label{eq:model_indepenedent_params}
    u=\sqrt{\frac{\mu_2}{\mu_1}},\quad
    k=\Delta \phi_{21},\quad
    z=z_2-z_1,
\end{equation}
where $\Delta \phi_{21}$ is the constant phase difference between the two images.  Initially, we use these model-independent parameters to avoid complications that can arise from the periodicity of the likelihood with respect to $M_l$ and $v$ and from the unboundedness of $y_0$. The numerical samples are then converted to $\{M_l,y_0,v\}$.

The lens and source parameters can be accurately measured, even at large impact parameter, despite degenerate posteriors. The  three top panels in Fig.~\ref{fig:comparison} show 68\% C.L.~limits on the lensing parameters from the 1-dimensional marginalized posteriors, obtained from the MC samples,
as a function of the initial impact parameter $y_0$ (fiducial setup). Since the mass of {Sgr A$^*$} is known with sub-percent accuracy \cite{Gravity2019}, we run two analyses: one with the lens mass $M_l$ treated as a free parameter, and another with $M_l$ fixed to its known value.  In both cases, the lens parameters can be extracted from the signal beyond $y_0=1$. For example, for $y_0=3$ all parameters are constrained with a relative error smaller than $50\%$ at $1\sigma$ C.L. 
    At large enough $y_0$, the sampled contours become consistent with $\mu_2=0$ (i.e., no second image detected) and lens parameters cannot be constrained. Note that, for strictly monochromatic sources, $M_l$ and $v$ only enter the GW signal phase and can be constrained only up to a periodic factor, see Eq.~\eqref{eq:amplifact}.
Figure  \ref{fig:triangleplot} shows the 1- and 2-d 95\% C.L.~marginalized lens parameter posteriors from the MC. For large impact parameter, particularly noticeable in the $y_0=3$ contour, the posteriors deviate from Gaussian behaviour. For impact parameter $y \gtrsim 4$, the lack of resolution of the small amplitude modulation, i.e.~the $r$ contour being compatible with $r=0$, prevents the reconstruction of the lensing parameters (cf. Fig.~\ref{fig:triangleplot_full} and  Fig.~\ref{fig:triangleplot_full_original} for the complete set of 1-d and 2-d marginalized 2-$\sigma$ posteriors).

To assess our capacity to differentiate between the lensed and null (i.e., unlensed) hypotheses, we compute the Bayes ratio between the two models, denoted as $\mathcal{B}_{LU}$.
The bottom panel in Fig.~\ref{fig:comparison} shows its variation as a function of the initial impact parameter for different distance ratios $r_{ls}/r_l$ (fiducial setup). Very strong evidence for the lensed signal (i.e.~$\log \mathcal{B}_{LU} >10$) can be robustly established for $y_0 \le y_{\rm max} \simeq 5 $ (2.2) for $r_{ls}/r_l = 0.1$ ($2$). A similar analysis for advanced LIGO shows a more modest gain, with $y_{\rm max} \simeq 1.5$, for $r_{ls}/r_l=1$. 
Note that the evidence for the unlensed and static lens hypotheses are identical, as they are both degenerate for a monochromatic signal. 

We explored different setups varying the fiducial parameters. In the Gaussian limit, the covariance matrix scales as the inverse of the ${\rm SNR}\sim\sqrt{T_{\rm obs}}\mathcal{A}/r_s$ (cf.~Eq.~\eqref{eq:snr}). For fixed SNR, changing the frequency of the source influences our results marginally, as long as the modulation period in Eq.~\eqref{eq:modulation_period} is smaller than or comparable to the observation time. Similarly, while parameter estimation requires a non-zero  $\veff$, the error and the evidence for the lensed signal is rather insensitive to its exact value as long as $\veff \gtrsim 10^{-5} \approx 3\, {\rm km/s}$, which is high enough to observe the modulation period. Thus, small velocities are unlikely to prevent the observation of lensed NS. Moreover, the sensitivity to very small velocities suggests that CWs could also measure lens accelerations. The parameter with the largest influence on the precision is  the distance ratio $r_{ls}/r_{l}$, with smaller values leading to better parameter estimation and larger evidence.


\subsection{Detection Prospects} \label{sec:probability_estimates}

We will now estimate the probability of strong lensing and compare it with previous results.
We will follow the source distribution proposed in Ref. \cite{Reed:2021scb} (also considered in Ref. \cite{Basak:2022fig}),
\begin{equation}\label{eq:ns_spatial_distrib}
    \frac{dP_s(r,z)}{dV} = \frac{1}{2\pi}\frac{1}{\sigma_r^2}e^{-\frac{r^2}{2\sigma_r^2}} \frac{1}{2\Delta z} e^{-\frac{|z|}{\Delta z}}\,.
\end{equation}
Here $dV =dr r d\phi dz$ is the volume element, $\sigma_r=5$kpc is the radial scatter of sources and $\Delta z$ is the scatter perpendicular to the galactic plane. We will assume that the Earth and any detectable lensed source deviate negligibly from the galactic plane, relative to $\Delta z\sim 0.1-1$kpc, and set $z\approx0$.

\begin{figure}
    \centering
    \includegraphics[width=\columnwidth]{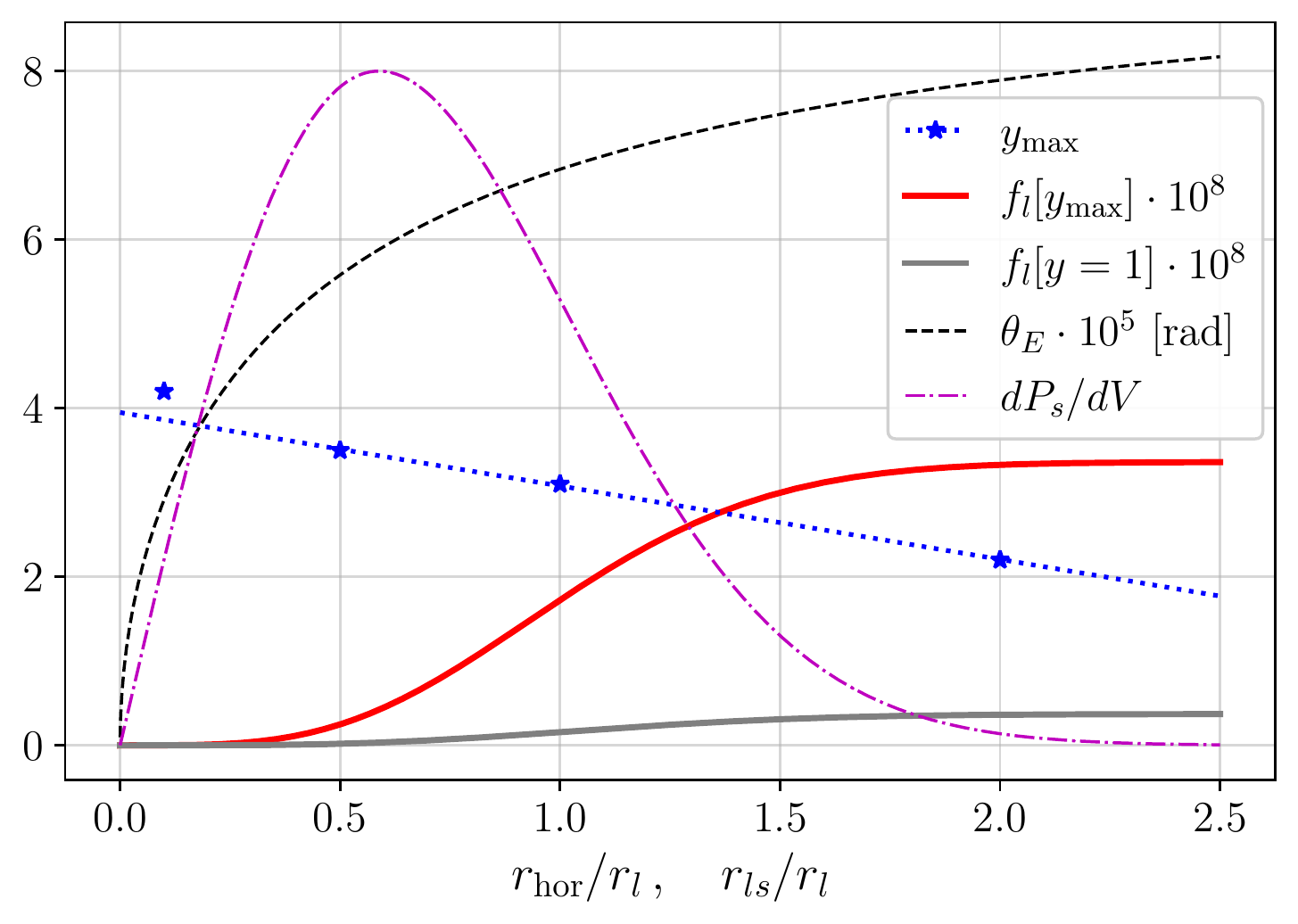}
    \caption{Expected fraction of NSs lensed by Sgr A$^*$ using the strong-evidence cutoff (thick solid red) and the Einstein radius cutoff (thick solid grey) as a function of $r_{\rm hor}/r_l-1$ for $\Delta z = 0.1$kpc. Thin lines represent other quantities appearing in Eq.~\eqref{eq:fraction_lensed_sources}, plotted as a function of $r_{ls}/r_l$: Einstein angle (dashed), source density (dash-dotted), values of $y_{\rm max}$ interpolated (dashed) from the results of the MC sampling (stars).}
    \label{fig:probability_estimates_factors}
\end{figure}

The fraction of strongly lensed sources is
\begin{equation}\label{eq:fraction_lensed_sources}
f_l\equiv \frac{\bar N_l}{N_0} \approx  \int_0^{r_{\rm hor}} d r_{ls} r_{ls}(r_{l}+r_{ls})\left(\theta_E y_{\rm max}\right)^2 \frac{dP_s}{dV}\,,
\end{equation}
were $\bar N_l$ is the average number of lensed sources and $N_0$ is the number of observable sources \cite{Basak:2022fig}. The integral is performed up to a detection horizon $r_{\rm hor}$ (cf.~Eq.~\ref{eq:NS_horizon}) and $y_{\rm max}$ is the highest value of the impact parameter for which very strong evidence for lensing can be established, see Fig.~\ref{fig:comparison}.
For $r_{ls}/r_l\geq 0.5$, $y_{\rm max}$ obeys a linear relation, which we have extrapolated to the whole domain to compute the integral above. This underestimates $y_{\rm max}$ close to the lens  (stars vs line in Fig.~\ref{fig:probability_estimates_factors}) and is thus a conservative assumption.

 Figure \ref{fig:probability_estimates_factors} shows the fraction of lensed NSs with strong evidence, $f_l[y_{\rm max}]$, as a function of $r_{\rm hor}$, where we also show the fraction of sources within the Einstein radius, ${f_l[y=1]}$, obtained from setting $y_{\rm max}\to 1$ in Eq.~\eqref{eq:fraction_lensed_sources}. The Einstein radius, source distribution and strong-evidence parameter entering the calculation are also shown.
For $r_{\rm hor}\gtrsim 1.5\,r_l$, the expected number of observed sources is
\begin{equation}
{\bar N_L}=3.36\left(\frac{0.1{\rm kpc}}{\Delta z}\right)\left(\frac{N_0}{10^8}\right)\,.
\end{equation}
This is a factor $\sim 9.04$ larger than the estimate obtained setting $y_{\rm max}=1$ in Eq.~\eqref{eq:fraction_lensed_sources}, as in the analysis of Ref.~\cite{Basak:2022fig}. 
The number of detected sources is given by Poisson statistics, with
\begin{equation}
    P_l(k) = \frac{\bar N_l^k}{k!}e^{-\bar N_l}\,,
\end{equation}
where $k$ is the number of detections. Hence, the probability of detecting at least a single lensed event is $P_l(k\geq 1)=1-e^{-N_l}$. The analysis in Ref.~\cite{Basak:2022fig} uses $N_0=10^9$ as a fiducial number of observable sources. In that case, and for a narrow spread $\Delta z=0.1$ kpc, our study suggests that an average of 34 strong-evidence lensed events would be observed, with $\sim 2\cdot 10^{-15}$ chance of observing none.

Situations that mimic strongly lensed sources can be ruled out. 
A source in a binary may exhibit an amplitude modulation via precession \cite{Apostolatos:1994mx,Breton:2008xy}. However, this can be distinguished from lensing through a periodic phase difference,  which is absent for lensed sources moving at constant velocity.

\begin{table}
\begin{tabular}{lcc}
\toprule
{} &   spherical &        disk \\ \hline
\newline
spatial   &   0.059 &     771.6\\
frequency &     \multicolumn{2}{c}{7.02} \\
both      & $4.1\cdot 10^{-9}$ & $5.4\cdot 10^{-5}$ \\\hline\hline
\end{tabular}
\caption{Expected number of overlapping signals that may mimic strong lensing signatures. All these quantities scale with the number of detectable NSs as $N_0/10^8$.
\label{tab:overlap_chances}}
\end{table}

Finally, two unrelated sources may be close enough in frequency and sky localization to appear as two images of a strongly lensed signal. However, the chances of such a coincidence are negligible: to mimic strong lensing, their angular separation must be $\lesssim 10''$ (cf.~Fig.~\ref{fig:resolving_images_perturber}) and their frequencies might differ by no more than $\Delta f/f \lesssim 10^{-7}( \alpha_{ij}/10'')(v/3 \times 10^{-3} )$, see Eq.~\eqref{eq:modulation_period}. 
Table \ref{tab:overlap_chances} gives the expected number of overlapping signals out of a total of $N_0=10^8$ detectable sources. For the spatial overlap, we consider sources randomly distributed in the sky, whose rate of overlap is $\sim N_0\delta\theta^2/4$, as well as sources confined to the galactic disk $\sim N_0 \theta/2\pi$ (both cases corresponds to the limit of large and small $\Delta z$ in  Eq.~\eqref{eq:ns_spatial_distrib} below). 
Regarding the frequency, we will consider sources distributed homogeneously in the range $\log(f_0)=[0.1,1]$ kHz, so the overlap rate is $\sim N_0\Delta f/(f\log(10))$. While the chance of spatial or frequency overlap is sizeable, the probability of both occurring simultaneously is negligible. 

Additional information can be further used to constrain this possibility, including the alignment of the images relative to Sgr A*, the signals' relative amplitude, and the frequency evolution.


\subsection{Angular resolution of GO images}
\label{sec:resolve_images}
CWs also enable accurate sky localization. In order to identify a CW in a blind search, it is necessary to ``undo'' the detector's motion by redefining the time variable: $t=t^{\prime} + \hat n\cdot \boldsymbol{x}(t^\prime) + \Delta T$ \cite{Brady:1997ji}. 
Here $t^\prime$ is the observation time, $\hat n$ is the source's direction, $\boldsymbol{x}$ is the detector's position relative to the Solar System barycenter and $\Delta T$ is the relativistic time delay. Both Earth's rotation and orbital motion contribute to $\boldsymbol{x}$, but the  latter becomes dominant after an integration period of a few days ~\cite{Maggiore:2007ulw}.
Once a signal is identified, analysis exploiting the complete coherence of the signal can accurately determine an image's position, depending on the source location. 

We use Eq.~(7.151) of Ref.~\cite{Maggiore:2007ulw} to estimate the sky localisation accuracy. This depends on the relative inclination between the Earth's orbital plane and the source direction, so the resulting skymap is not isotropic. We compute the mismatch between the two waveforms $(\delta h| \delta h)$ (cf. Eq.~\eqref{eq:innerprodf}). In our case, $\delta h$ is the difference between two monochromatic waveforms resampled with different source directions, $({\hat n}, {\hat n}' )$, so
\begin{equation}\label{eq:mm_sky}
    (\delta h| \delta h)\simeq 2 \rho^2 \left[ 1 - \frac{1}{T}\int_0^T dt\cos{\left(2 \pi f_0 \boldsymbol{x}(t) \cdot  ({\hat n}- {\hat n}')\right)} \right].
\end{equation}
Imposing the condition $(\delta h| \delta h)=1$ returns the sky localization variance \cite{Lindblom:2008cm}, $\sigma_\theta$, as a function of the sky orientation of the vector ${\hat n}- {\hat n}'$.

\begin{figure}[t!]
    \centering
    \includegraphics[width=1\columnwidth]{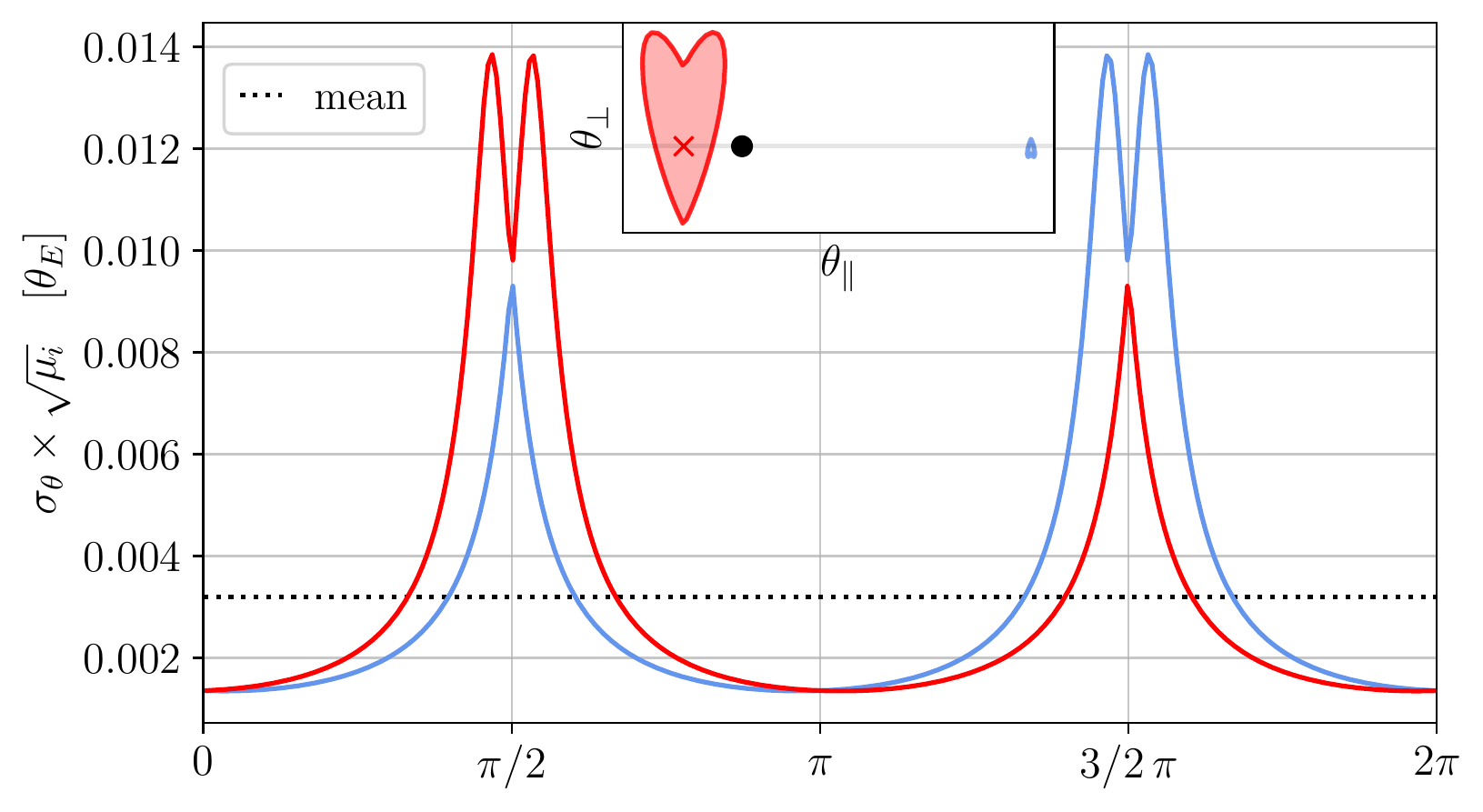}
    \caption{
    Accuracy in resolving image sky positions. The figure displays the 1-$\sigma$ angular precision for determining the sky position of images resulting from CWs sources lensed by Sgr A$^*$. The blue curve corresponds to the brighter image, while the red curve to the fainter image. Both are rescaled by the respective magnification factor. The resulting skymaps are shown in the inset for illustrative purposes.
    }
    \label{fig:sky_map}
\end{figure}

In the context of lensed sources, the accuracy of determining their sky localization varies by the SNR of the observed images, $\rho_i=\sqrt{\mu_i}\rho$. Fig. \ref{fig:sky_map} shows the 1-$\sigma$ angular accuracy on the image sky position that can be achieved for a CW signal lensed by SgrA$^*$ for our fiducial setup. The shape of the skymap is a squeezed cardioid that is rescaled and flipped (because of spatial parity) between the two images. We employed the module Astropy \cite{astropy:2022} to model Earth's orbital motion relative to Sgr A$^*$ position in Eq.\eqref{eq:mm_sky}.  Sgr A$^*$ is put at a right ascension of 17h 45m 40.0409s and a declination of -29$^{\circ}$ 0' 28.118" \cite{Reid:2004rd}.

In the inset of Fig.~\ref{fig:resolving_images_perturber}, the heart-shaped maps are tilted with respect to the optical axis by an angle $\phi_{\rm sky}$ that encodes the source-galactic plane relative inclination, and their size scales as $\propto 1/(\sqrt{\mu_i}{\rm SNR})$.

A simple analytical estimate can be derived considering the Earth to be on a circular orbit around the Sun, with radius $1\,{\rm AU}$, and the source direction on the orbital plane.
Then, an observation time of $T_{\rm obs}>1\,{\rm yr}$ corresponds to an angular accuracy of \begin{equation}
    {\sigma_\theta\simeq 21\,{\rm mas} \left(60/{\rm SNR}\right)\left(1\,{\rm kHz}/ f_0\right)}\,.
\end{equation}
This is comparable to the mean angular accuracy of the anisotropic setup.

In general, we expect to be able to individually resolve the images for lenses with mass $M_l> 1.2 \times 10^3 M_\odot \left(r_l / 8.18\,{\rm kpc}\right)\left(r_{ls}/r_s\right)\left(60/{\rm SNR}\right)\left(1\,{\rm kHz}/ f_0\right)$, at $y=1$.


\section{Probing Sgr A$^*$ companions}\label{sec:probing}
Let us consider the prospect of searching for additional objects near the main lens via their effect on the image positions. A smoking gun for such objects is a misalignment between Sgr A$^*$ and the two lensing images, as this would require breaking the axial symmetry. We first derive the perturbation to the lensing observables (Sec. \ref{sec:app_perturbations}) induced by a perturber and discuss when its presence can be probed through image misalignment (Sec. \ref{sec:misall}.

\subsection{Effects of a perturber on GO images}\label{sec:app_perturbations}

We first study how the presence of a point-like perturber near the lens affects the lensing observables. Let us introduce the dimensionless lensing potential, the so-called Fermat potential, 
\begin{equation}
T(\boldsymbol{x})=\frac{t(\boldsymbol{x})}{4GM_l}\;,
\end{equation}
and consider the effect of a point lens perturber with mass $M_{\rm pert} = m \cdot M_{\rm Sgr A^*},$ at $\boldsymbol{x}_m=(x_m^\parallel,x_m^\perp)$ (here $\boldsymbol{x}$ is defined by normalizing to the Einstein angle of the unperturbed lensed, as defined above).
The total lensing potential is
\begin{equation}\label{eq:fermat_perturbed}
    T(\boldsymbol{x}) = T^{(0)}(\boldsymbol{x}) + m\log(|\boldsymbol{x}-\boldsymbol{x}_m|)\,,
\end{equation}
where $T^{(0)}(\boldsymbol{x})$ is the unperturbed Fermat potential.

We will assume $m\ll1$ sufficiently small for a perturbative treatment and $y$ large enough for the source to be outside the caustic network of the binary lens. Under our assumptions, the system forms an additional image at 
\begin{equation}
\boldsymbol{x}_3\approx \boldsymbol{x}_m - \boldsymbol{M}^{-1} \cdot \boldsymbol{\Delta}_{ym} \;,
\end{equation}
where the vector $\boldsymbol{\Delta}_{ym}$ is defined as
\begin{equation}
    \boldsymbol{\Delta}_{ym} = \boldsymbol{y}-\boldsymbol{x}_m\left(1-x_m^{-2}\right)\;,
\end{equation}
and the elements of the matrix $\boldsymbol{M}$ are given by
\begin{equation}
    M_{ij} = \delta_{ij} \bigg( \frac{\Delta^2_{ym}}{m} + \frac{1}{ x_m^2} \bigg) - \frac{2 {x}_{m}^i {x}_m^j}{x_m^4} \;.
\end{equation}

The third image will undergo a frequency shift and induce an additional amplitude modulation, similar to the interference effect described above. However, the amplitude of the third image is typically very small, as $\mu_3\propto m^2/|\boldsymbol{y}-\boldsymbol{x}_m|^4$. Even for GWs, $\sqrt{|\mu_3|}h_0$ will most likely be below the detection threshold. Moreover, the existence of an additional image at arbitrary $y$ is a feature of point lenses and other very compact matter distributions,  while it is absent in generic extended lenses (cf.~Ref.~\cite{Tambalo:2022wlm}, Sec IIIA). We will thus focus on the perturber's effect on the position and time delay between the main images.  (Given that $\delta\mu_\pm/\mu_\pm \propto m/|\boldsymbol{y}-\boldsymbol{x}_m|^2$, the effect on the magnification is negligible in most cases.)

The leading-order effect on the Fermat potential of each image is
\begin{equation}
 T_\pm - T_\pm^{(0)}= m\log(|\boldsymbol{x}_\pm-\boldsymbol{x}_m|)\,,
\end{equation}
where we have used Fermat's principle on the unperturbed lens, i.e.~$\boldsymbol{\nabla} T_{\pm}^{(0)} = 0$. (A more accurate result follows from evaluating the full Fermat potential \eqref{eq:fermat_perturbed} on the perturbed image positions, Eqs.~(\ref{eq:Delta_x_parallel}, \ref{eq:Delta_x_perp}).)  
Hence, the time delay between the images is proportional to
\begin{equation}
    \Delta T_\pm \approx m\log\left(\frac{|\boldsymbol{x}_+-\boldsymbol{x}_m|}{|\boldsymbol{x}_--\boldsymbol{x}_m|}\right)\,.
\end{equation}
Restoring the lens' mass gives us the effect of a perturber on the relative phase  between the GO images:
\begin{equation}
   k = 8\pi G M_l f_0 \Delta T_\pm \sim  5.14\cdot 10^5 \frac{m}{x_m} \left(\frac{f_0}{{\rm kHz}}\right) \,.
\end{equation}

Therefore, a perturber with mass $m$ can affect the phase at observable levels up to $x_m\lesssim 8\cdot 10^4 m/\sigma_k$, where $\sigma_k\sim 0.3$ (cf. Fig. \ref{fig:triangleplot_full_original}) is the sensitivity to the phase. The expected number of such objects scales as the projected density times $x_m^2$ and it's likely to be significant: a perturber with $m\sim 2\cdot 10^{-6}$ (corresponding to a mass $\sim 10M_\odot$, for which $8\pi G M_l f\sim 1$, i.e.~at the WO diffraction limit for a $1$kHz source) can influence the signal beyond the Einstein radius. 
Nevertheless, the actual sensitivity will be much degraded by the degeneracies between the lens mass and motion parameters (velocity, acceleration), cf.~Fig.~\ref{fig:triangleplot}. Therefore, the contribution of light objects to the time delay will result in a slight shift of these parameters posteriors, with relative magnitude $\sim \sum_i m_i\ll 1$.


\subsection{Finding perturbers through image misalignment}\label{sec:misall}

\begin{figure}[t!]
    \centering
    \includegraphics[width=1\columnwidth]{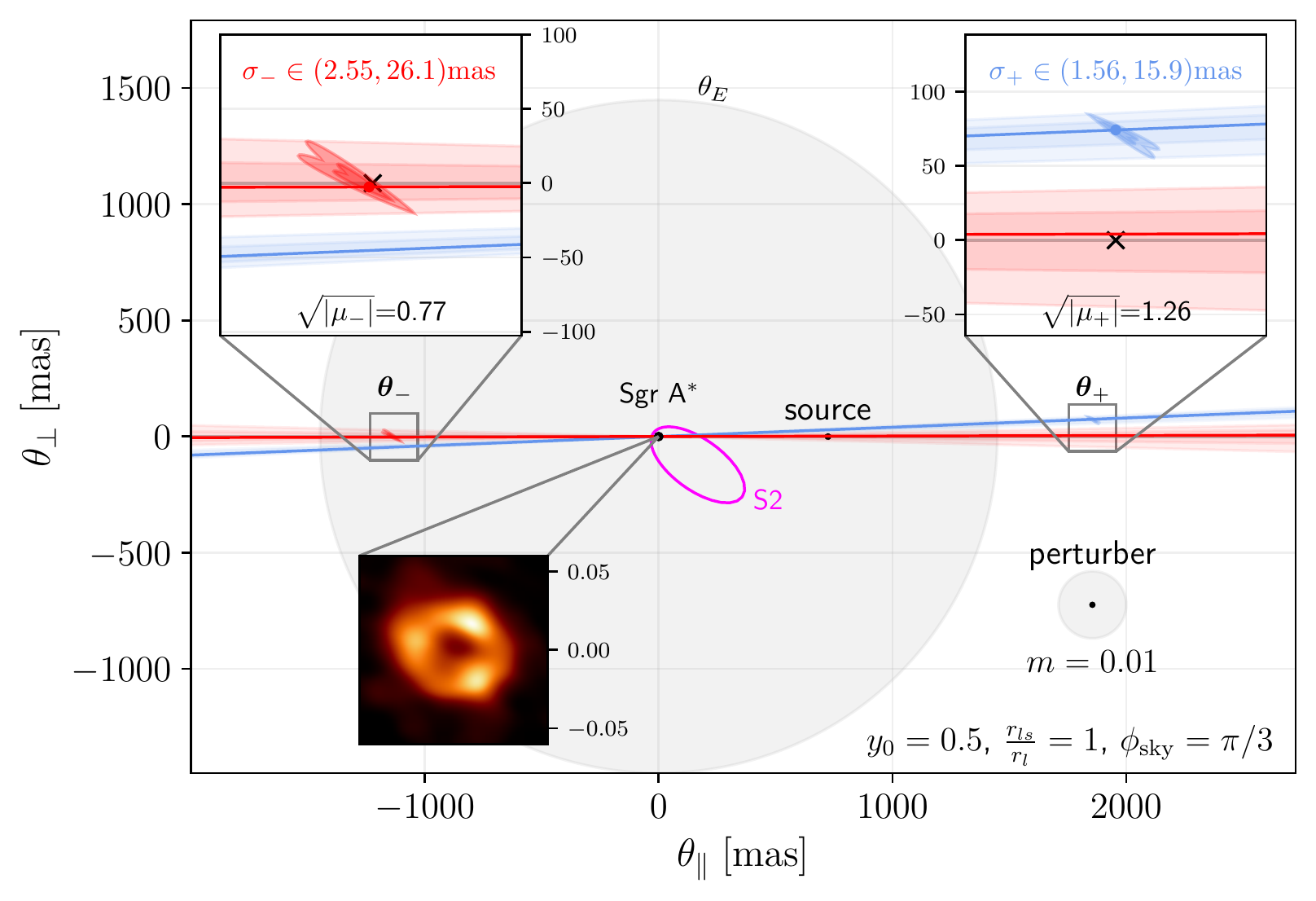}
    \caption{Resolving individual images. The insets show the 68,95\% C.L.~localization regions (filled shapes), the unpertubed images (crosses) and the projected lens-image axes (filled bands). 
    A $\sim 4\cdot 10^4M_\odot$ perturber at $\boldsymbol{x}_m=(1.5,-0.5)$ displaces the positive-parity image by $\gtrsim\sim 3\sigma$  away from the optical axis. Einstein radii are shown in gray. The projected orbit of the S2 star \cite{GRAVITY:2020gka} and the image of Sgr A$^*$ \cite{EventHorizonTelescope:2022wkp} are shown for comparison. }
    \label{fig:resolving_images_perturber} 
\end{figure}

\begin{figure}
    \includegraphics[width=\columnwidth]{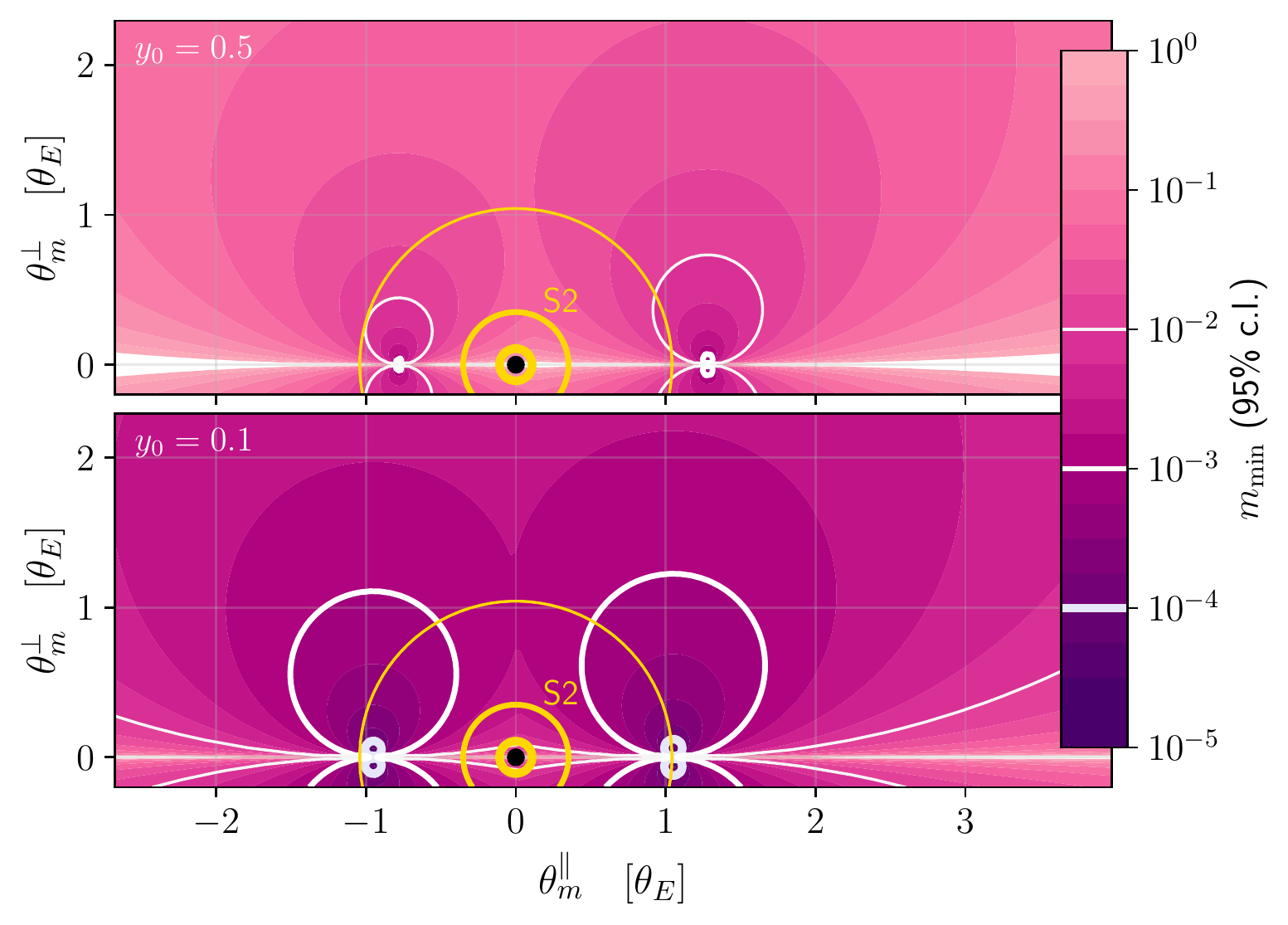}
    \caption{Minimum perturber mass producing a detectable lens-image misalignment (2-$\sigma$, cf.~Fig.~\ref{fig:resolving_images_perturber}), depending on its position in the lens plane. The source is our fiducial setup at $y=0.5,0.1$ (upper/lower), assuming the median error in sky localization. Lines indicate $10\%, 1\%$ and $0.1\%$ of the mass of Sgr A$^*$. Orange lines centered around Sgr A$^*$ show the corresponding 95\% C.L.~limits from the Schwarzschild precession of the star S2 \cite{GRAVITY:2020gka}, assuming $r_{ls}=r_l$.}
    \label{fig:perturber_m_min}
\end{figure}

Let us now turn to the effects of a perturber on the image positions.
In the absence of perturber, the positive/negative parity images lie in the optical axis (i.e. the source-lens axis) at $\boldsymbol{x}_\pm=(x_\pm,0)$. The perturber displaces the main images by 
\begin{align}\label{eq:Delta_x_parallel}
    \Delta x^\parallel_\pm &\approx \frac{m}{1+x_\pm^{-2}}\frac{x_\pm-{x}_m^{\parallel}}{{(\boldsymbol{x}_\pm - \boldsymbol{x}_m)}^2} \,, 
    \\ \label{eq:Delta_x_perp}
 \Delta x^\perp_\pm &\approx \frac{-m}{1-x_i^{-2}}\frac{{x}_m^{\parallel}}{{(\boldsymbol{x}_\pm - \boldsymbol{x}_m)}^2}\,, 
\end{align}
The image displacement along the optical axis \eqref{eq:Delta_x_parallel} can be probed, but it is degenerate with the main lens parameters ($y$ and $M_l,r_s$ via $x_E$). In contrast, the off-axis displacement \eqref{eq:Delta_x_perp} is a smoking gun for additional structure.

A perturber will generically produce an off-axis displacement of the images, $\Delta x_\pm^\perp\neq0$. Because the optical axis is not known, one can only measure a misalignment between the two images and the black hole. If the sky localization uncertainty were isotropic, the uncertainty in the $\pm$ image displacement with respect to the optical axis is given by
\begin{equation}
\sigma^{\perp}_{\pm} = \sqrt{\sigma_\pm^2 + ({x_\mp}/{x_\pm})\sigma_\mp^2} = \sigma_\pm \sqrt{1+\left({x_\pm}/{x_\mp}\right)^3}\,.
\end{equation}
The first equality follows from adding in quadrature the uncertainty on the image and the projected uncertainty of the other image along the direction of the lens (see Fig. \ref{fig:resolving_images_perturber}). The second equality employs the scaling of an image's localization accuracy with the magnification, $\sigma_\pm^2=\sigma_\mp^2 \mu_\mp/\mu_\pm$, and the fact that for a point lens $\mu_-/\mu_+=x_-/x_+$. Precision degrades at large $y$ because the negative parity image becomes faint (poorly localized) and close to Sgr A$^*$, thus reducing the level arm. 

To account for the anisotropic sky-localization uncertainty $\boldsymbol{\sigma}_\pm$ (see discussion in Sec. \ref{sec:resolve_images}), we consider the projected uncertainty in the image's axis orientation (through the angle with respect to Sgr A$^*$) as $\varphi_\pm(\varphi, \phi_{\rm sky}) = \arctan\left(\frac{(\boldsymbol{x}_\pm + \boldsymbol{\sigma}_{\pm}(\phi_{\rm sky}) )_2}{(\boldsymbol{x}_\pm + \boldsymbol{\sigma}_{\pm}(\phi_{\rm sky}) )_1}\right)$. Here, $\varphi$ is the angle that defines the orientation of a vector $\boldsymbol{\theta}=(\theta_1,\theta_2)$ in the lens plane, and $\phi_{\rm sky}$ is the angle that set the orientation of the optical axis with respect to the Earth's orbital plane and of the shape of the anistoropic sky-map.
Then the uncertainty in the projection of the axes is given by $\sigma_\pm^{\perp} \in \left(\min_{\varphi} (\phi_\pm), \max_{\varphi} (\phi_\pm)\right) x_\pm$.
In practice the $+$ image is always better localized (both because $\mu_+>|\mu_-|$ and $x_+>|x_-|$). Therefore, we can interpret the uncertainty $\sigma_+^{\perp}$ as associated with the optical axis (i.e. measurement of $\phi_{\rm sky}$). The (larger) uncertainty in the negative-parity image $\sigma_-^{\perp}$ gives the sensitivity to deviations from axial symmetry and the presence of perturbers. 
Because uncertainties depend strongly on the fiducial value of $\phi_{\rm sky}$, we will quote the median sensitivities. Note, however, that the sensitivity improves substantially for certain configurations, e.g. $\phi_{\rm sky}\sim \pm \pi/2$ (this information is available from the localization of the positive-parity image). Combining sky-localization and time-domain analysis will further improve the sensitivity.


Following the sky localization accuracy discussed above, each image defines an axis through the known lens position. 
Figure \ref{fig:resolving_images_perturber} illustrates this setup with an example perturber leading to a $\sim3\sigma$ measurable displacement. The blue/red shaded regions represent the 1-2$\sigma$ confidence bands on the lens-image axis.
In practice, the axis of the positive parity image $x_+$ determined more precisely due to better intrinsic localization and larger level arm (i.e.~both $|\mu_+/\mu_-|,|\theta_+/\theta_-|>0$). Therefore, we consider $\boldsymbol{\theta}_+$ to fix the optical axis  and its uncertainty, while $\boldsymbol{\theta}_-$ determines the sensitivity to off-axis image displacement.

Figure \ref{fig:perturber_m_min} shows the 95\% C.L.~minimum detectable mass via the offset between Sgr A$^*$ and the two images of a rotating NS at $y=0.5,0.1$ (with fiducial parameters), as a function of the perturber's position. Precision degrades at large $y$, both because the negative parity image becomes faint (poorly localized) and close to Sgr A$^*$. For $y\lesssim 1$ the area over which a perturber can be detected scales as $\sim \frac{m}{y^3}$, increasing greatly for closely aligned systems. This analysis is conservative, as combining sky localization with time-domain information (cf. Fig. \ref{fig:comparison}) will improve sensitivity.
This method is complementary to other probes of the galactic center: analogue constraints form pericenter passage of the S2 star \cite{GRAVITY:2020gka} probe the region near the star's orbit, rather than the GO images. Thus, lensed NSs are sensitive to different regions in the galactic center. In addition, gravitational lensing is sensitive to perturbers at intermediate distances (i.e. between the observer and lens or between the lens and source). 

\vspace{5pt}
\section{Discussion and prospects}
Lensed CWs offer strong complementarity to lensed EM signals and GWs from compact binary coalescences, bringing new challenges and opportunities. We have established the prospect of reconstructing the lens parameters and further probing the lens by individually resolving the images. We have focused on rotating NSs lensed by Sgr A$^*$, but our conclusions extend to other systems.

The prospect of identifying lens systems at several Einstein radii leads to enhanced detection probabilities. The number of potentially lensed sources scales with the square of $y_{\rm max}$, i.e.~the maximum impact parameter within which a detection can be established with strong evidence. In Ref.~\cite{Basak:2022fig}, it was shown that 3G detectors can detect up to $\sim 6$ NSs within the Einstein radius of Sgr A$^*$, corresponding to $y_{\rm max}=1$. Even when applying a stringent threshold for strong evidence (cf.~Fig.~\ref{fig:comparison}), we find that $y_{\rm max}$ can be as large as $\sim 3$. This increases in the number of detectable sources by an order of magnitude compared to the estimate based on $y_{\rm max}=1$.
While the ultimate prospects depend on unknown astrophysics, a targeted search of NSs lensed by Sgr A$^*$  is warranted. 

Individual images can be resolved with $\sim 10$mas accuracy (Fig.~\ref{fig:resolving_images_perturber}), comparable to the best optical telescopes. 
As an application, we showed how lens-image misalignment provides a smoking gun for additional structure (Fig. \ref{fig:perturber_m_min}), probing companion objects in regions complementary to stellar orbits \cite{Naoz:2019sjx,GRAVITY:2021xju,Will:2023nlt} and BH imaging \cite{EventHorizonTelescope:2022wkp}.
Future work will address additional signatures of these objects and the benefits of combining timing and sky-localization information.

CWs lensed by Sgr A$^*$ provide a novel probe of the galactic center.
Detection of lensed EM radiation from objects closely aligned with Sgr A$^*$ is challenging due to abundant stars, gas and dust in the central region of the galaxy \cite{Bozza:2012by}. In contrast, lensed CWs are negligibly absorbed, providing a pristine view of the region near our supermassive BH. CWs will complement other approaches to probe the matter distribution around Sgr A*, test dark matter scenarios \cite{Gondolo:1999ef,Hui:2016ltb,DeLuca:2023laa} and GW propagation \cite{Ezquiaga:2020dao,Goyal:2023uvm,Oancea:2022szu,Oancea:2023hgu,Eichhorn:2023iab} in an extreme environment.

This study is a first step towards understanding parameter reconstruction in lensed CWs.
We have focused on NSs lensed by Sgr A$^*$, but our results apply to other lenses and sources. Lensing of CWs from NSs in all-sky searches will probe compact objects and galactic substructure, enabling novel tests of intermediate-mass black holes and the dark matter distribution. 
Our treatment can be extended to other long-lived sources, such as inspiraling compact binaries observable by planned or proposed space detectors ~\cite{LISA:2017pwj,Gong:2021gvw,Sedda:2019uro,Baibhav:2019rsa}. Expanding beyond our simplifying assumptions---i.e.~constant magnifications, geometric optics and point-like lenses--- will unveil the full potential of lensed CWs and enable novel tests of astrophysics and fundamental physics.

\begin{acknowledgments}
We are very grateful to Soummyadip Basak,  Jing Ming, Maria Alessandra Papa, Ornella Juliana Piccinni, Giovanni Tambalo and Hector Villarrubia-Rojo for discussions, Lorenzo Speri, Jonathan Gair and Ollie Burke for help with aspects of the parameter estimation, and the anonymous referees for valuable and constructive criticism.
\end{acknowledgments}

\appendix


\section{Parameter inference}\label{app:parinfer}

The signal observed at the detector, $d(t)$, is a superposition between background noise and the true gravitational waves signal,
\begin{equation}\label{eq:data}
    d (t)= h(t;\boldsymbol{\theta}) + n(t).
\end{equation}
 The vector $\boldsymbol{\theta}$ includes the parameters that determine the propagated waveform.
For a stationary and Gaussian distributed noise, a realisation $n_0$ has a probability
\begin{equation}
    p(n_0)\propto \exp \left[-\frac{1}{2}\int df \frac{|\Tilde{n}_o(f)|^2}{S_n(f)}\right]\,,
\end{equation}
where the one-sided power spectral density of noise, $S_n(f)$, is the variance associated with the noise distribution. 

The probability of observing $d$ given $\boldsymbol{\theta}$, i.e.~the \emph{likelihood} of the parameters $\boldsymbol{\theta}$, is
\begin{equation}\label{eq:whittle_likelihood}
    \log p(d \mid \boldsymbol{\theta}) \propto -\frac{1}{2}(d - h(\boldsymbol{\theta}), d - h(\boldsymbol{\theta}))\,,
\end{equation}
where  the inner product for two signals $f$ and $g$, in the Fourier space, is defined as
\begin{equation}\label{eq:innerprodf}
    (f,g)=4\,\text{Re}\left[\int_0^\infty df\, \frac{\Tilde{h}^{\star}(f)\,\Tilde{g}(f)}{S_n(f)}\right]\,.
\end{equation}
For quasi-monochromatic sources, the inner product in Eq.~\eqref{eq:innerprodf} can be equivalently defined in the time-domain, following Ref.~\cite{Takahashi:2002ky, Seto:2002dz}, as
\begin{equation}\label{eq:innerprodt}
(h,g) = \frac{2}{S_{n}(f_{0})}\int_{0}^{T_{\mathrm obs}}h(t)g(t)\,,
\end{equation}
with $T_{\mathrm obs}$ the observational time. Hence, the SNR, which measures the loudness of the signal, reads
\be
\rho \equiv (h|h) \simeq \sqrt{ \frac{\mathcal{A}^2 }{r_s^2} \frac{T_{\rm obs}}{ S_n(f_0)} } \; .
\ee

The posterior of $\boldsymbol{\theta}$ is obtained from the likelihood using Bayes’ theorem, i.e.,
\begin{equation}
    p(\boldsymbol{\theta}\mid d)=\frac{p(d \mid \boldsymbol{\theta})p(\boldsymbol{\theta})}{p(d|\mathcal{M}_i)} \;,
\end{equation}
where $p
(\boldsymbol{\theta})$ is the parameter prior and $p(d|\mathcal{M}_i)$, namely the \emph{evidence}, is the marginal likelihood for a given model $\mathcal{M}_i$:
\begin{equation}
p(d|\mathcal{M}_i)= \int d\boldsymbol{\theta} p(d|\boldsymbol{\theta})p(\boldsymbol{\theta})\,.
\end{equation}
In the Bayesian context, the marginalized posteriors are the probability distribution functions of the parameters.

Given two models $\mathcal{M}_i$ and $\mathcal{M}_j$, the ratio of their evidence called the \textit{Bayes factor}, is used as an index to test different interpretations of the same data:
\begin{equation}
 \mathcal{B}_{ij} =\frac{p(d|\mathcal{M}_i)}{p(d|\mathcal{M}_j)}\,.
\end{equation}
Following the empirical Jeffrey’s prescription \cite{Kass:1995loi}, in Tab.~\ref{tab:bayes_ratio} we report the range of value of $ \mathcal{B}_{ij}$ and the corresponding degree of evidence of the model $\mathcal{M}_i$ over $\mathcal{M}_j$.
\newline
\begin{table}[!h]
\begin{center}
\begin{tabular}{||p{2cm}  p{2cm}  p{3.5cm}||} 
 \hline
$\log\mathcal{B}_{10}$ & $\mathcal{B}_{10}$ & {Evidence against $H_0$}  \\ [0.5ex] 
 \hline\hline
 2 to 6 & 3 to 20 & Positive   \\ 
 \hline
 6 to 10 & 20 to 150 & Strong  \\
 \hline
 $>$10 & $>$150 & Very Strong  \\ 
 \hline
\end{tabular}
\caption{\label{tab:bayes_ratio}Evidence against the null hypothesis compared to ranges of Bayes factor.}
\end{center}
\end{table}


\begin{figure*}
    \centering
  \includegraphics[width =\textwidth]{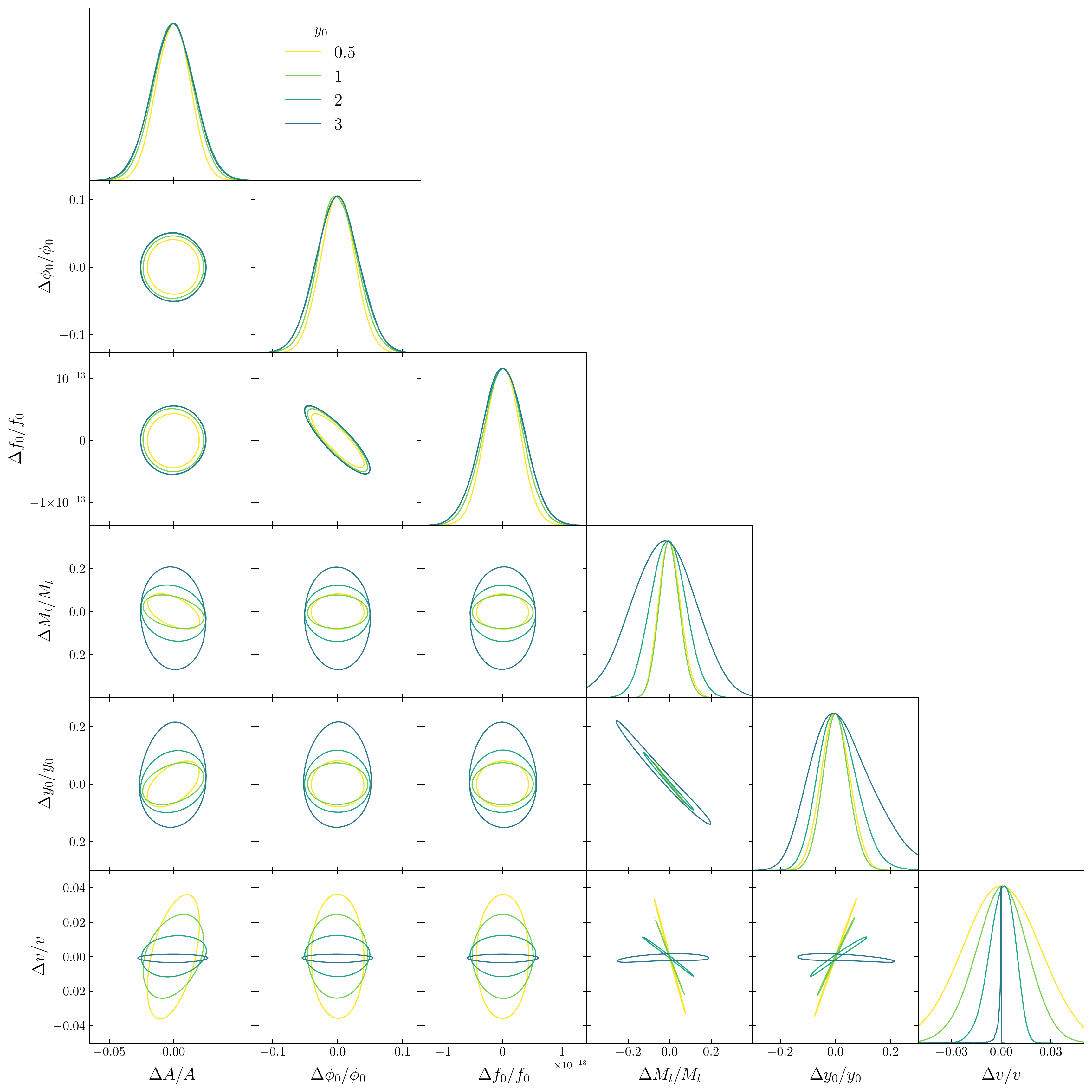}
  \caption{1-d and 2-d marginalized 2-$\sigma$ posteriors of the full set of source and lens parameters, for the fiducial setup discussed in the text. Here, $A \equiv \mathcal{A}/r_s$.}
  \label{fig:triangleplot_full}
\end{figure*}

\begin{figure*}
    \centering
    \includegraphics[width =\textwidth]{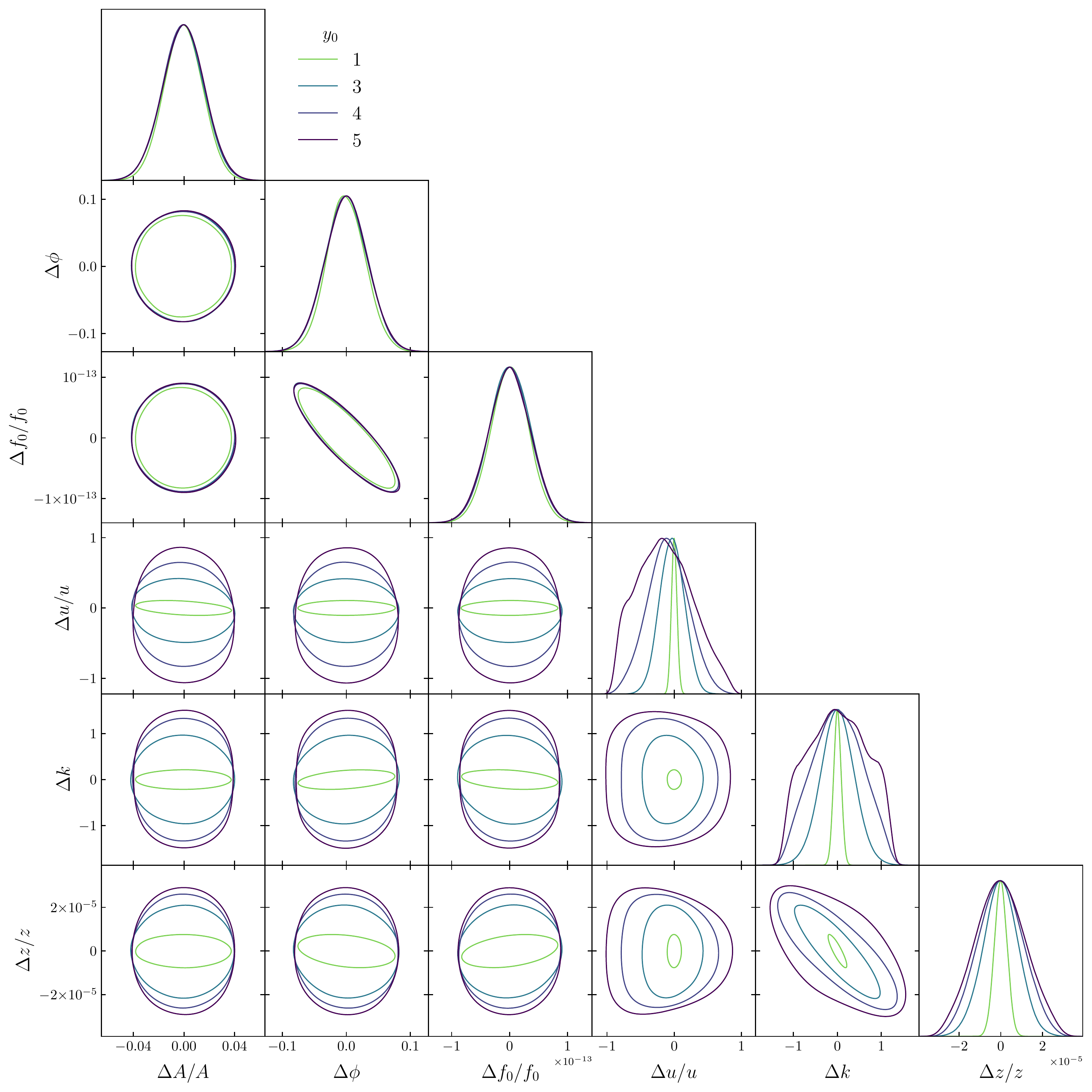}
    \caption{1-d and 2-d marginalized 2-$\sigma$ posteriors of the full set of source and lens parameters, with the latter expressed in terms of the original sampling parameters,  for the fiducial setup discussed in the text. Here, $A \equiv \mathcal{A}/r_s$.}
    \label{fig:triangleplot_full_original}
\end{figure*}

\end{document}